\documentclass[prb,twocolumn,superscriptaddress,showpacs]{revtex4}

\usepackage{graphicx}
\usepackage{dcolumn}
\usepackage{amsmath}
\usepackage{color}

\newcommand{\BFA}{BaFe$_{2}$As$_{2}$}
\newcommand{\BKFA}{Ba$_{1-x}$K$_x$Fe$_{2}$As$_{2}$}

\begin{document}
\title{Magnetic fluctuations and superconductivity in Fe pnictides\\ probed by Electron Spin Resonance}

\author{N.~Pascher}
\author{J.~Deisenhofer}
\author{H.-A.~Krug von Nidda}
\author{M.~Hemmida}
\affiliation{EP V, Center for Electronic Correlations and Magnetism,
Institute for Physics, Augsburg University, D-86135 Augsburg,
Germany}

\author{H. S. Jeevan}
\author{P. Gegenwart}
\affiliation{I. Physik. Institut, Georg-August-Universit\"{a}t
G\"{o}ttingen, D-37077 G\"{o}ttingen, Germany }

\author{A.~Loidl}
\affiliation{EP V, Center for Electronic Correlations and Magnetism,
Institute for Physics, Augsburg University, D-86135 Augsburg,
Germany}

\date{\today}

\begin{abstract}
The electron spin resonance absorption spectrum of Eu$^{2+}$ ions
serves as a probe of the normal and superconducting state in
Eu$_{0.5}$K$_{0.5}$Fe$_2$As$_2$. The spin-lattice relaxation rate
$1/T_1^{\rm ESR}$ obtained from the ESR linewidth exhibits a
Korringa-like linear increase with increasing temperature above
$T_{\rm c}$ evidencing a normal Fermi-liquid behavior. Below 45~K
deviations from the Korringa-law occur which are ascribed to
enhanced magnetic fluctuations within the FeAs layers upon
approaching the superconducting transition. Below $T_{\rm c}$ the
spin lattice relaxation rate $1/T_1^{\rm ESR}$ follows a
$T^{1.5}$-behavior without any appearance of a coherence peak.

\end{abstract}



\maketitle

\section{Introduction}

The recent discovery of superconductivity in Fe-based pnictides and
chalcogenides\cite{Kamihara08,Chen08,Rotter08a,Hsu08} has triggered
enormous research efforts to understand the origin of
superconductivity and its relation to the inherent magnetism of
iron. One class of these materials are the ternary
\textit{A}Fe$_2$As$_2$ systems with \textit{A} = Ba, Sr, Ca, Eu
(122-systems) and $T_{\rm c}$ values up to
38~K.\cite{Rotter08a,Jeevan08b,Sefat08,Leithe-Jasper08} The parent
compounds exhibit a spin-density wave (SDW) anomaly accompanied by a
structural distortion.\cite{Rotter08b,Rotter09} Superconductivity (SC)
appears, e.g., by substituting the $A$-site ions by
K\cite{Rotter08a,Jeevan08b} or Fe by
Co.\cite{Sefat08,Leithe-Jasper08} For underdoped \BKFA{} and Co
doped \BFA{} a coexistence of the SDW state and superconductivity
has been
reported.\cite{Rotter09,Rotter08-Angewandte,Chen09,Pratt09,Christianson09,Kant09}

A particularly interesting 122-system is EuFe$_2$As$_2$ with
$T_{\rm{SDW}}$ = 190~K,\cite{Raffius93,Jeevan08a,Wu09} the highest
reported SDW transition temperature in the pnictides. This system is
of special importance among the 122 iron-pnictides, since the
antiferromagnetic ordering of local Eu$^{2+}$ moments at $T_{\rm N}=19$~K
provides the opportunity to study the interplay between Eu and Fe
magnetism and also the influence of Eu magnetism on SC (under
hydrostatic pressure~\cite{Miclea09} or
doping~\cite{Jeevan08b,Zheng09,Ren09}). The appearance of a SDW gap
in EuFe$_2$As$_2$ was evidenced by optical spectroscopy\cite{Wu09}
and, recently, the opening of even two gaps with different
characteristics was reported.\cite{Moon09} Moreover, electron spin
resonance (ESR) in single crystalline EuFe$_2$As$_2$ revealed a
drastic change in the magnetic properties of the Eu-spin system from
a typical metallic-like behavior above $T_{\rm{SDW}}$ to a behavior
characteristic for a magnetic and insulating system in the SDW
state.\cite{Dengler09}

Here, we focus on Eu$_{0.5}$K$_{0.5}$Fe$_2$As$_2$ in which the iron
SDW is completely suppressed by hole doping and SC is found below
$T_{\rm c} = 32$~K.~\cite{Jeevan08b} The bulk nature of SC is confirmed by
a clear specific-heat anomaly and diamagnetism found in DC-magnetization
and AC-susceptibility measurements. After
subtraction of the phonon contribution, a specific-heat jump height
of about 70~mJ/molK$^2$ has been deduced.~\cite{Jeevan09}
M\"{o}ssbauer spectroscopy measurements have established the
coexistence of Eu$^{2+}$ short-range magnetic ordering with SC in
Eu$_{0.5}$K$_{0.5}$Fe$_2$As$_2$ below 4.5~K.~\cite{Anupam09} At the
same temperature, a peak is found in the zero-field cooled
magnetization, measured at low fields of 5~mT,~\cite{Jeevan08b} and
a corresponding minimum occurs in the magnetic penetration depth
$\lambda(T)$, determined by radio-frequency
technique.~\cite{Gasparov} Recently, the substitution of Fe by Co in
EuFe$_2$As$_2$ reportedly leads to an incomplete superconducting
transition in the electrical resistivity.\cite{Zheng09} SC has also
been found in chemically pressurized
EuFe$_2$(As$_{0.7}$P$_{0.3}$)$_2$ at $T_{\rm c}=26$~K, followed by
ferromagnetic Eu-ordering at 20 K.\cite{Ren09}

In this work we show that the ESR signal of Eu$^{2+}$ can be used as
probe of the superconducting properties in the 122-family of Fe
pnictides. In a metallic system the linewidth of the ESR absorption
is a direct measure of the spin-lattice relaxation rate $1/T_1^{\rm
ESR}$, thus providing information on the density of states at the
Fermi energy and the opening of the SC gap. In polycrystalline
Eu$_{0.5}$K$_{0.5}$Fe$_2$As$_2$ we observe a clear change of
$1/T_1^{\rm ESR}$ from a normal Fermi-liquid like behavior with a
Korringa relaxation $\propto T$ above 45~K, the onset of magnetic
fluctuations  of the FeAs layers for $T_{\rm c} < T < 45$~K, and a
$\propto T^{1.5}$-law below $T_{\rm c}$ = 32~K.

\section{Experimental details}

Polycrystalline Eu$_{0.5}$K$_{0.5}$Fe$_2$As$_2$ was prepared using a
sintering method described in Ref.~\onlinecite{Jeevan08b} and
characterized by energy-dispersive X-ray (EDX) analysis, x-ray,
electrical resistivity, magnetic susceptibility, and specific-heat
experiments.~\cite{Jeevan09} ESR measurements were performed in a
Bruker ELEXSYS E500 CW-spectrometer at X-band frequencies ($\nu
\approx 9.36$~GHz) equipped with a continuous He gas-flow cryostat
in the temperature region $4.2 < T< 300$~K. ESR detects the power
$P$ absorbed by the sample from the transverse magnetic microwave
field as a function of the static magnetic field $H$. The
signal-to-noise ratio of the spectra is improved by recording the
derivative $dP/dH$ using lock-in technique with field modulation.
The sample was measured in the form of fine powder.

\section{Experimental Results and Discussion}

\begin{figure}
\centering
\includegraphics[width=70mm,clip]{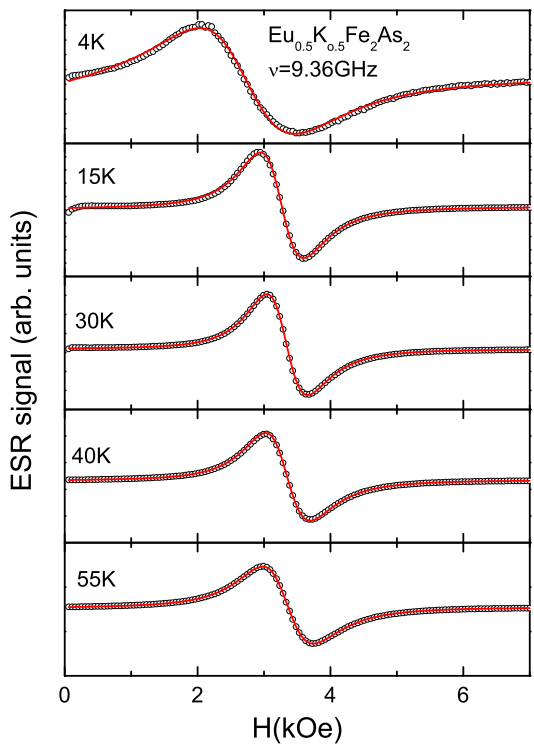}
\vspace{2mm} \caption[]{\label{spectra} (Color online) ESR spectra
and corresponding fit curves of Eu$_{0.5}$K$_{0.5}$Fe$_2$As$_2$ at
different temperatures below and above $T_{\rm c}$.}
\end{figure}

\begin{figure}
\centering
\includegraphics[width=75mm,clip]{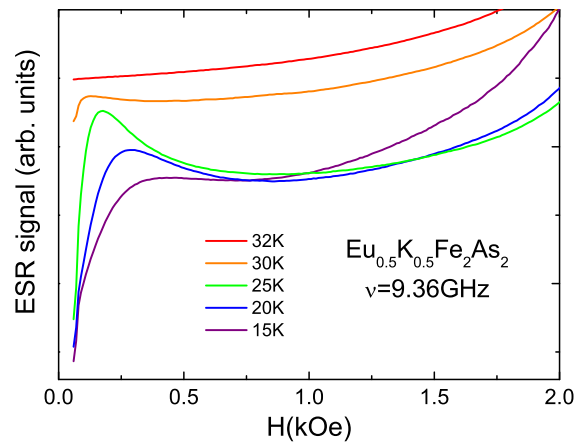}
\vspace{2mm} \caption[]{\label{spectralowfield} (Color online)
Temperature evolution of the low-field absorption spectra of
Eu$_{0.5}$K$_{0.5}$Fe$_2$As$_2$ below $T_{\rm c}$. Lines are shifted
for clarity.}
\end{figure}

Figure \ref{spectra} shows ESR spectra of a powdered polycrystal of
Eu$_{0.5}$K$_{0.5}$Fe$_2$As$_2$ for different temperatures. In all
cases one observes a single exchange-narrowed resonance line which
is well described by a Dyson shape,\cite{Barnes1981} i.e. a Lorentz
line at resonance field $H_{\rm res}$ with half width at half
maximum $\Delta H$ and a contribution of the dispersion to
absorption $(D/A)$ ratio $0 \leq D/A \leq 1$, resulting in an
asymmetry typical for metals where the skin effect drives electric
and magnetic components of the microwave field out of phase. The
$D/A$ ratio is determined as a fit parameter and depends on sample
size, geometry, and skin depth. If the skin depth is small compared
to the sample size, $D/A$ approaches 1.  Focussing on the low-field
regime, below the superconducting transition temperature $T_{\rm c}$
the spectra as documented in Fig.~\ref{spectralowfield} exhibit a
dip-like signal at low fields typical for the magnetic shielding
below the lower critical field $H_{\rm c1} \approx
100$~Oe,\cite{Gasparov} followed by a broad non-resonant microwave
absorption feature due to the penetration of magnetic flux in the
Shubnikov phase above $H_{\rm c1}$ of the type II
superconductor.\cite{Blazey1987}

Returning to the resonant absorption we determined the absolute
value of the ESR spin susceptibility $\chi_{\rm ESR}$ above $T_{\rm
c}$ by comparison of the double-integrated signal intensity with
that of the reference compound Gd$_2$BaCuO$_5$ -- the so called
green phase -- which exhibits an ESR signal with a similar
linewidth. In Gd$_2$BaCuO$_5$ all Gd$^{3+}$ spins (with the same
electron configuration $4f^7$ and spin $S=7/2$ like Eu$^{2+}$)
contribute to the ESR signal and the corresponding susceptibility
exhibits a Curie-Weiss law with $\Theta_{\rm
CW}=-23$~K.\cite{Goya1996} We find that about 50\% of the Eu$^{2+}$
ions participate in the resonant absorption. Estimating the skin
depth $\delta=(\rho/\mu_0\omega)^{0.5}$ using the resistivity value
$\rho =$~0.04~$\Omega$~cm at room temperature\cite{Anupam09} and the
microwave frequency 9 GHz we find a skin depth of $\delta\approx$
75~$\mu$m. The typical grain size of powdered samples is of the same
order of magnitude in agreement with the percentage of Eu$^{2+}$
spins contributing to the observed signal.

\begin{figure}[t]
\centering
\includegraphics[width=80mm]{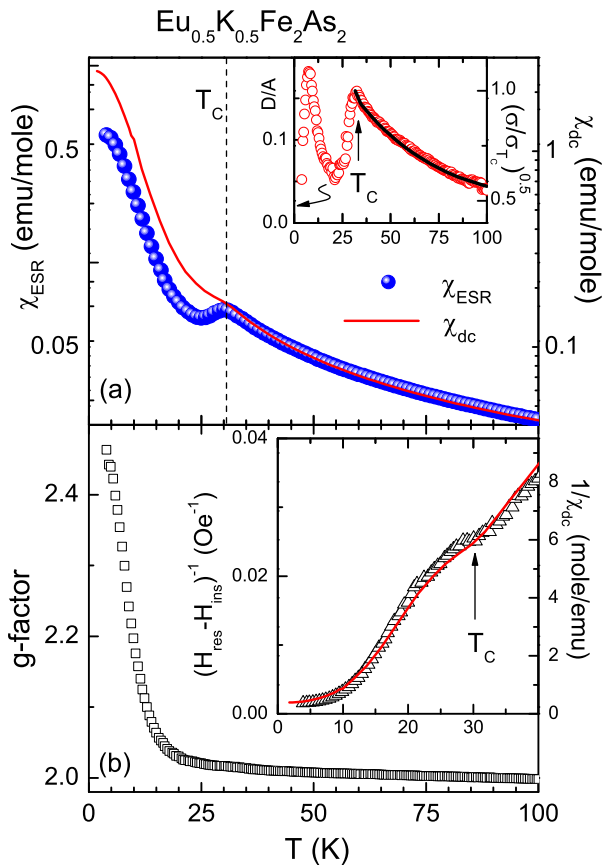}
\vspace{2mm} \caption[]{\label{int} (Color online) (a) Temperature
dependence of the ESR intensity compared to the magnetic DC
susceptibility (solid line) measured in a magnetic field of 0.3~T,
i.e. well above the lower critical field $H_{\rm c1}$ in
Eu$_{0.5}$K$_{0.5}$Fe$_2$As$_2$. Inset: Temperature dependence of
the dispersion to absorption ratio $D/A$ compared to the square root
of the normalized DC conductivity (solid line). (b): Temperature
dependence of the $g$-factor of the ESR line. Inset: Inverse shift
of the resonance field $H_{\rm res}$ from its insulator value
$H_{\rm ins}$ given by $g_{\rm ins}=1.993$ compared to the inverse
static susceptibility.}\end{figure}

\begin{figure}[t]
\centering
\includegraphics[width=80mm]{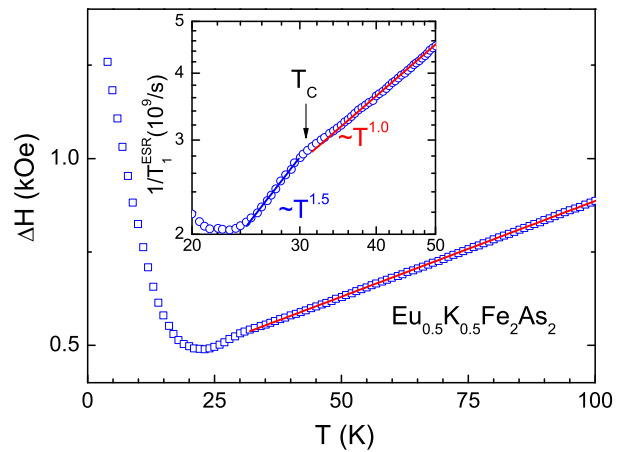}
\vspace{2mm} \caption[]{\label{dh} (Color online) Temperature
dependence of the ESR linewidth $\Delta H$. The solid line indicates
the linear Korringa law. Inset: Temperature dependence of
$1/T_1^{\rm ESR}$ on a double-logarithmic scale and linear fits
indicating the power-laws in the normal and SC states.}
\end{figure}

The temperature dependence of the double-integrated signal intensity
$\chi_{\rm ESR}$ is compared to the static susceptibility $\chi_{\rm
dc}$ in Fig.~\ref{int}(a). Note the different scales due to the fact
that only 50\% of the Eu spins contribute to the resonant
absorption. Starting from high temperatures the Curie-Weiss behavior
of $\chi_{\rm ESR}$ is interrupted just below $T_{\rm c}$, where it
abruptly decreases and again increases on further lowering the
temperature, while the static susceptibility increases monotonously
without any drop  but only slightly deviating from the Curie-Weiss
law at $T_{\rm c}$. The temperature dependence of the $D/A$ ratio is
shown in the inset of Fig.~\ref{int}(a) together with the square
root of the normalized conductivity $(\sigma/\sigma(T=T_{\rm
c}))^{0.5}$ in the normal state taken from
Ref.~\onlinecite{Jeevan08b}. In the normal state the two quantities
can be scaled to fall on top of each other, confirming that $D/A$ is
proportional to the penetration depth of the microwave. The $D/A$
ratio drops by a factor of 3 when crossing $T_{\rm c}$ with
decreasing temperature. After the anticipated onset of enhanced
magnetic fluctuations of the Eu ions at about 25~K it starts to
increase again up to 8~K, below which $D/A$ decreases again. The
drop of intensity on passing $T_{\rm c}$ can be understood due to
the fact that magnetic resonance is observed only from the volume
fraction of the sample which is penetrated by the magnetic flux,
i.e. the surface within the London penetration depth and -- in
superconductors of type II -- the magnetic flux tubes with normal
conductivity.  The concomitant drop of the $D/A$ ratio is difficult
to explain, because the superconductivity strongly reduces the  skin
depth and, hence, is naively expected to increase the $D/A$ ratio.
The observed decrease may presumably result from the change of the
effective geometry of the normal-state regions, i.e. the flux tubes
in the superconducting matrix instead of a homogeneously conducting
state, because the conductivity of these regions can be anticipated
to be unchanged. Thus, the ratio of the skin depth to the diameter
of the flux tubes is larger than the ratio of the skin depth to the
full sample diameter, which can result in a decrease of the D/A
ratio. However, detailed electro dynamical considerations are
necessary to clarify this observation. The consecutive increase of
the $D/A$ ratio on decreasing temperature can be related to the
reentrant Eu magnetism which leads to an increase of the normally
conducting volume fraction. Finally, the peak at about 8~K marks the
onset of short-range order.\cite{Gasparov,Anupam09}

The temperature dependence of the $g$ value depicted in the lower
frame of Fig.~\ref{int} is only slightly affected by the onset of
superconductivity. The $g$-value is about 2 above $T_{\rm c}$ and
starts to increase strongly below 25~K. The inset compares the
corresponding inverse shift of the resonance field from its value in
an insulating environment determined by $g_{\rm ins}=1.993$ (see
Ref.~\onlinecite{Abragam1970}) to the inverse static susceptibility.
Both quantities approximately coincide showing that the resonance
shift is dominated by demagnetization fields resulting from the
large Eu magnetization similar to observations in systems like
GdI$_2$ or YBaMn$_2$O$_6$.\cite{Deisenhofer2004,Zakharov08}
 Only the small kink at $T_{\rm c}$ in both the
reciprocal susceptibility and the resonance shift can be regarded as
an effect of the superconducting state due to the opening of the
excitation gap in the electronic density of states at the
Fermi-level and corresponding reduction of the Pauli contribution to
the susceptibility.

The $g$ shift at elevated temperature $\Delta g = g-g_{\rm ins} =
J_{\rm CE-Eu}(0) N(E_{\rm F}) \approx 0.02$ results from the
homogenous polarization of the conduction electrons in the external
field (Pauli susceptibility) and is comparable to usual
metals.\cite{Taylor1975}

Now we will turn to the temperature dependence of the ESR linewidth
shown in Fig.~\ref{dh}. Note that in metallic systems the ESR
linewidth is determined by the spin-lattice relaxation time $T_1$
\cite{Barnes1981} and, hence, provides  information complementary to
NMR or NQR measurements. One can clearly identify a linear increase
with temperature with a slope $b = 5.1$~Oe/K and a residual
zero-temperature width $\Delta H_0$ = 374~Oe. Upon entering the
superconducting state, a pronounced drop of the linewidth can be
recognized in the inset of Fig.~\ref{dh}. Below about 20~K the
linewidth increases again due to growing magnetic fluctuations of
the Eu spins on approaching short-range order.

The observed linear increase of the linewidth $\Delta H \propto
\langle J_{\rm CE-Eu}^2(q) \rangle N^2(E_{\rm F}) T$ (Korringa
relaxation) is a typical signature of local moments in a Fermi
liquid and depends on the conduction-electron density of states
$N(E_{\rm F})$ at the Fermi energy $E_{\rm F}$ and the exchange
coupling $J_{\rm CE-Eu}$ between the conduction electrons and the Eu
spins. The observed slope of 5.1~Oe/K is a usual value for S-state
$4f^7$ local moments in metals and indicates a normal
three-dimensional Fermi-liquid state.\cite{Taylor1975, Barnes1981,
Elschner1997} In case of ferromagnetic correlations between the Eu
ions (reading of the Curie-Weiss temperature from the inset in the
lower panel of Fig.~\ref{int} gives $\Theta_{CW}\simeq 10$K) the
residual linewidth is expected to be negative $\Delta H_0 = -
b\Theta_{\rm CW}\simeq -50$~Oe with $b$ being the derived Korringa
slope.\cite{Barnes1981}  The larger positive value $\Delta H_0$ =
374~Oe is probably due to an inhomogeneous distribution of the Eu
and K ions, a problem which is well-known for the 122-systems. This
may lead to strong fluctuations of the long-range dipolar fields
while the narrowing effect of the short-range exchange narrowing is
reduced.

\begin{figure}[t]
\centering
\includegraphics[width=80mm]{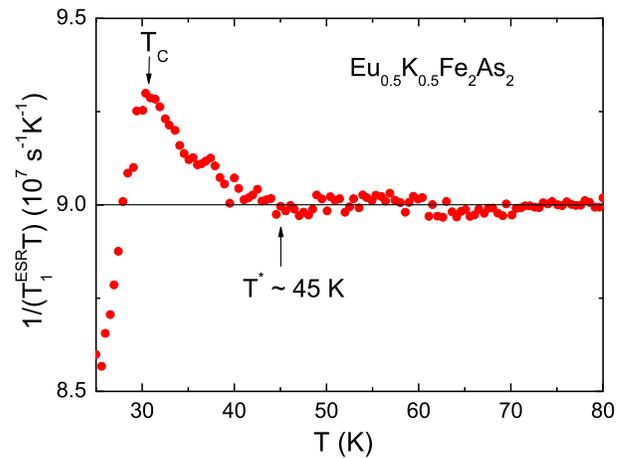}
\vspace{2mm} \caption[]{\label{fig:T_one} (Color online) Temperature
dependence of $1/(T_1^{\rm ESR}T)$. The solid line indicates the
linear Korringa law. Below $T^*\approx$ 45~K a deviation from the
linear behavior indicates the onset of magnetic fluctuations on
approaching the superconducting transition.}
\end{figure}

In the inset of Fig.~\ref{dh} we show the temperature dependence of
the reciprocal electron spin-lattice relaxation time given by
\begin{equation}
1/T_1^{\rm ESR}=\gamma(\Delta H-\Delta H_{0}),
\end{equation}
where $\Delta H_{0}$ = 374 Oe is the residual zero-temperature
linewidth obtained from the linear fit in the normal regime and
$\gamma$ denotes the gyromagnetic ratio. The obtained power-law in
the normal state clearly confirms the Korringa-law, but for the
superconducting state we observe a behavior $\propto T^{1.5}$
without any indication of a coherence (Hebel-Slichter) peak, ruling
out a conventional BCS scenario with an isotropic gap. Due to the
onset of magnetic fluctuations the ESR power-law can only be traced
in a narrow temperature range below $T_{\rm c}$  and we cannot
exclude the possibility that $1/T_1^{\rm ESR}$ may be influenced by
the vicinity of the short-range Eu interactions. To examine closer
the vicinity of $T_{\rm c}$ we plot $1/(T_1^{\rm ESR}T)$ as a
function of temperature in Fig.~\ref{fig:T_one}. A clear increase
from the constant high-temperature (Korringa) value occurs already
below a temperature $T^*\simeq 45$~K and leads to a maximum and a
sharp drop just below $T_{\rm c}$. A similar behavior of the
spin-lattice relaxation time of $^{57}$Fe and $^{75}$As above
$T_{\rm c}$ has been reported by NMR in the related system
Ba$_{0.6}$K$_{0.4}$Fe$_{2}$As$_{2}$ and attributed to spin
fluctuations in the FeAs layers due to interband
nesting.\cite{Yashima09}

A Korringa behavior in the normal state has also been found by ESR
in pure and Co-doped EuFe$_2$As$_2$ \cite{Dengler09,Ying09} and by
NMR and NQR studies on other
122-compounds,\cite{Fukazawa09,Yashima09} but in the superconducting
state of Fe-based superconductors the temperature behavior of the
nuclear spin-lattice relaxation time has revealed non-universal
power-laws $1/T_1\propto T^\alpha$ with $\alpha$ ranging from
2.5-6.\cite{Parker08,Fukazawa09,Yashima09,Kobayashi09} We want to
point out that an NQR study of pure KFe$_2$As$_2$ revealed a
power-law $\propto T^{1.4}$ similar to our ESR results which was
interpreted in terms of multi-gap superconductivity with line
nodes,\cite{Fukazawa09} and a recent NMR study suggested that this
exponent may be universal for overdoped
Ba$_{1-x}$K$_x$Fe$_{2}$As$_{2}$. \cite{Zhang10}

Given the narrow temperature range of the ESR power-law, we refrain
at the present stage from performing a fit of the data, as the above
mentioned NMR studies showed that very different power-laws can be
obtained by varying the gaps' sizes and symmetry. However, our data
nicely show that local-moment ESR in iron pnictides is a highly
efficient tool to shed light on the superconducting order parameters
as a complementary method to nuclear spin resonance techniques.

\section{Summary}

In summary, we show that the ESR signal of Eu$^{2+}$ spins gives
direct access to the superconducting state of the 122-class of
pnictides. We identify a normal Fermi-liquid behavior above $T_{\rm c}$
from the Korringa-law of the ESR spin-lattice relaxation rate
$1/T_1^{\rm ESR}$. Just above $T_{\rm c}$ we observe a deviation from the
Korringa behavior which we assign to magnetic fluctuations in the FeAs layers on
approaching the superconducting state. Below $T_{\rm c}$ no Hebel-Slichter
peak is observed, ruling out a simple isotropic BCS scenario, and
the spin-lattice relaxation rate follows $1/T_1^{\rm ESR}\propto T^{1.5}$.

\begin{acknowledgments}
We thank S. Graser for fruitful discussions and A. Pimenova and V.
Tsurkan for experimental support. We acknowledge partial support by
the Deutsche Forschungsgemeinschaft (DFG) via the Schwerpunktprogramm
SPP1458, the Collaborative Research Center TRR 80 and the
Research Unit FOR 960 (Quantum phase transitions).
\end{acknowledgments}


\begin{thebibliography}{39}

\bibitem{Kamihara08} Y. Kamihara, T. Watanabe, M. Hirano, and H. Hosono, J. Am. Chem.
Soc. \textbf{130}, 3296 (2008).

\bibitem{Chen08} X. H. Chen et al., Nature \textbf{453}, 761(2008).

\bibitem{Rotter08a} M. Rotter, M. Tegel and D. Johrendt, Phys.~Rev.~Lett. {\bf 101}, 107006 (2008).


\bibitem{Hsu08} F.-C. Hsu, J.-Y. Luo, K.-W. Yeh, T.-K. Chen, T.-W.
Huang, P. M. Wu, Y.-C Lee, Y.-L. Huang, Y.-Y. Chu, D.-C. Yan, and
M.-K. Wu, Proc. Natl. Acad. Sci. U.S.A. \textbf{105}, 14263 (2008).

\bibitem{Jeevan08b} H. S. Jeevan, Z. Hossain, D. Kasinathan, H. Rosner, C.
Geibel and P. Gegenwart, Phys. Rev. B \textbf{78}, 092406 (2008).

\bibitem{Sefat08} A. S. Sefat, R. Jin, M. A. McGuire, B. C. Sales, D. J. Singh, and D.
Mandrus, Phys. Rev. Lett. \textbf{101}, 117004 (2008).


\bibitem{Leithe-Jasper08} A. Leithe-Jasper, W. Schnelle, C. Geibel, and H. Rosner, Phys. Rev. Lett. \textbf{101}, 207004
(2008).


\bibitem{Rotter08b} M. Rotter, M. Tegel, D. Johrendt, I. Schellenberg,
W. Hermes and R. P\"{o}ttgen, Phys. Rev. B \textbf{78} 020503
(2008).

\bibitem{Rotter09} M. Rotter, M. Tegel, I. Schellenberg, F.M.
Schappacher, R. P\"{o}ttgen,  J. Deisenhofer, A. G\"{u}nther, F.
Schrettle, A. Loidl, and D. Johrendt, New J. Phys. \textbf{11},
025014 (2009).

\bibitem{Rotter08-Angewandte}
M. Rotter, M. Pangerl, M. Tegel, D. Johrendt, Angew. Chem. Int. Ed.
\textbf{47}, 7949-7952 (2008).


\bibitem{Chen09} H. Chen, Y. Ren, Y. Qiu, Wei Bao, R. H. Liu, G. Wu, T. Wu, Y. L.
Xie, X. F. Wang, Q. Huang, X. H. Chen, Europhys. Lett. \textbf{85},
17006 (2009).

\bibitem{Pratt09} D. K. Pratt, W. Tian, A. Kreyssig, J. L. Zarestky, S. Nandi, N. Ni, S. L. Bud'ko, P. C. Canfield, A. I. Goldman, and R. J. McQueeney,
Phys. Rev. Lett. \textbf{103}, 087001 (2009).

\bibitem{Christianson09} A. D. Christianson, M. D. Lumsden, S. E. Nagler, G. J. MacDougall, M. A. McGuire, A. S. Sefat, R. Jin, B. C. Sales, and D.
Mandrus, Phys. Rev. Lett. \textbf{103}, 087002 (2009).

\bibitem{Kant09} Ch. Kant, J. Deisenhofer, A. G\"{u}nther, F. Schrettle, A. Loidl, M. Rotter, D. Johrendt, Phys. Rev. B \textbf{81}, 014529 (2010).

\bibitem{Raffius93} H. Raffius, M. M¨orsen, B. D. Mosel, W. M\"{u}ller-Warmuth, W.
Jeitschko, L. Terb\"{u}chte, and T. Vomhof, J. Phys. Chem. Solids
\textbf{54}, 135 (1993).

\bibitem{Jeevan08a} H. S. Jeevan, Z. Hossain, D. Kasinathan, H. Rosner, C. Geibel and P.
Gegenwart, Phys. Rev. B \textbf{78}, 052502 (2008).

\bibitem{Wu09} D. Wu, N. Barišic, N. Drichko, S. Kaiser, A. Faridian, M. Dressel,
S. Jiang, Z. Ren, L. J. Li, G. H. Cao, Z. A. Xu, H. S. Jeevan and P.
Gegenwart, Phys. Rev. B \textbf{79}, 155103 (2009).


\bibitem{Miclea09} C. F. Miclea, M. Nicklas, H. S. Jeevan, D. Kasinathan, Z. Hossain,
H. Rosner, P. Gegenwart, C. Geibel, F. Steglich, Phys. Rev. B
\textbf{79}, 212509 (2009).



\bibitem{Ren09} Z. Ren, Q. Tao, S. Jiang, C. Feng, C. Wang, J. Dai,
G. Cao, Z. Xu, Phys. Rev. Lett. \textbf{102}  137002 (2009).

\bibitem{Zheng09} Q. J. Zheng, Y. He, T. Wu, G. Wu, H. Chen, J. J. Ying, R. H. Liu, X. F. Wang, Y. L. Xie, Y. J. Yan, Q. J. Li, X. H. Chen, arXiv:0907.5547
(2009).



\bibitem{Moon09} S. J. Moon, J. H. Shin, D. Parker, W. S. Choi, I. I. Mazin, Y. S. Lee, J. Y. Kim, N. H. Sung,
        B. K. Cho, S. H. Khim, J. S. Kim, K. H. Kim, and T. W. Noh, Phys. Rev. B \textbf{81}, 205114 (2010).


\bibitem{Dengler09} E. Dengler, J. Deisenhofer, H.-A. Krug von Nidda, Seunghyun Khim,
J.S. Kim, Kee Hoon Kim, F. Casper, C. Felser, and A. Loidl, Phys.
Rev. B \textbf{81}, 024406 (2010).


\bibitem{Jeevan09} H. S. Jeevan and P. Gegenwart, J. Phys.: Conf. Ser. \textbf{200},
012060 (2010).

\bibitem{Anupam09} Anupam, P. L. Paulose, H. S. Jeevan, C. Geibel and Z. Hossain, J. Phys. Condens. Matter {\bf21}
265701 (2009).

\bibitem{Gasparov} V. A.Gasparov, H. S. Jeevan, P. Gegenwart, JETP Letters {\bf89} (2009)
343.

\bibitem{Barnes1981} S. E. Barnes, Adv. Phys. \textbf{30}, 801
(1981).

\bibitem{Blazey1987}  K. W. Blazey, K. A. M\"uller, J. G. Bednorz, W. Berlinger,
G. Amoretti, E. Buluggiu, A. Vera, and F. C. Matacotta, Phys. Rev. B
{\bf 36}, 7241 (1987).

\bibitem{Goya1996} G. F. Goya, R. C. Mercader, L. B. Steren, R. D. Sanchez, M. T.
Causa, and M. Tovar, J. Phys.: Condens. Matter \textbf{8}, 4529 (1996).

\bibitem{Abragam1970} A. Abragam and B. Bleaney, {\it Electron
Paramagnetic Resonance of Transition Ions}, Clarendon Press, Oxford
(1970).



\bibitem{Deisenhofer2004} J. Deisenhofer, H.-A. Krug von Nidda, A. Loidl,
K. Ahn, R. K. Kremer, and A. Simon, Phys. Rev. B \textbf{69}, 104407
(2004).

\bibitem{Zakharov08} D.V. Zakharov, J. Deisenhofer, H.-A. Krug von Nidda, A. Loidl, T.
Nakajima, and Y. Ueda, Phys. Rev. B \textbf{78}, 235105 (2008).


\bibitem{Taylor1975} R.H. Taylor, Adv. Phys. \textbf{24}, 681
(1975).

\bibitem{Elschner1997} B. Elschner and A. Loidl:
in: {\itshape Handbook on the Physics and Chemistry of Rare Earth},
ed. by K. A. Gschneidner (Jr.), L. Eyring, Elsevier Science B. V.,
Amsterdam, Vol. {\bf 24}, 221 (1997).




\bibitem{Yashima09} M. Yashima, H. Nishimura, H. Mukuda, Y. Kitaoka, K. Miyazawa, P. M. Shirage, K. Kiho, H. Kito, H. Eisaki, A. Iyo, J. Phys. Soc. Jap.
\textbf{78}, 103702 (2009).


\bibitem{Ying09} J. J. Ying, T. Wu, Q. J. Zheng, Y. He, G. Wu, Q. J. Li, Y. J. Yan,
Y. L. Xie, R. H. Liu, X. F. Wang and X. H. Chen, Phys. Rev. B
\textbf{81}, 052503 (2010).

\bibitem{Fukazawa09} H. Fukazawa, Y. Yamada, K. Kondo, T. Saito, Y. Kohori, K. Kuga, Y.
Matsumoto, S. Nabaksuji, J. Kito, P. M. Shirage, K.Kihou, N.
Takeshita, C.-H.- Lee, A. Iyo, H. Eisaki, J. Phys. Soc. Jap.
\textbf{78}, 083712 (2009).




\bibitem{Parker08} D. Parker, O. V. Dolgov, M. M. Korshunov, A. A. Golubov, and I. I.
Mazin, Phys. Rev. B \textbf{78}, 134524 (2008).

\bibitem{Kobayashi09} Y. Kobayashi, A. Kawabata, S. C. Lee, T. Moyoshi, M.
Sato, J. Phys. Soc. Jpn. \textbf{78}, 073704 (2009).

 \bibitem{Zhang10} S. W. Zhang, L. Ma, Y. D. Hou, J. Zhang, T.-L. Xia, G. F.
Chen, J. P. Hu, G. M. Luke, and W. Yu, Phys. Rev. B \textbf{81},
012503 (2010).



\end{thebibliography}
\end{document}